\documentclass{bioinfo}
\usepackage{amsfonts}
\copyrightyear{2010} \pubyear{}
\usepackage{amssymb}
\usepackage{url}
\usepackage{mathrsfs}
\usepackage{color}
\newcommand{\ssm}{\fontsize{7pt}{\baselineskip}\selectfont}

\usepackage{eucal}
\usepackage{eufrak}
\usepackage{afterpage}
\firstpage{1}

\theoremstyle{remark}

\makeatletter \@addtoreset{equation}{section}

\title[RNA-RNA interaction prediction based on multiple
sequence alignments]
{RNA-RNA interaction prediction based on multiple
sequence alignments}

\author[A.X.\ Li, M.\ Marz, J.\ Qin, C.M.\ Reidys]{%
  Andrew X. Li\,$^1$, Manja Marz\,$^2$,
  Jing Qin\,$^3$,
  Christian M.\ Reidys\,$^{1,4}$\footnote{to
    whom correspondence should be addressed.
    Phone: *86-22-2350-6800;
    Fax:   *86-22-2350-9272;
    \texttt{duck@santafe.edu}%
  }}

\address{%
  $^{1}$Center for Combinatorics, LPMC-TJKLC, Nankai University
        Tianjin 300071, P.R.~China\\
  $^2$ RNA Bioinformatics Group, Philipps-University Marburg,
    Marbacher Weg 6, 34037 Marburg, Germany\\
  $^{3}$Max Planck Institute for Mathematics in the Sciences,
    Inselstrasse 22, D-04103 Leipzig, Germany\\
  $^{4}$College of Life Science, Nankai University
         Tianjin 300071, P.R.~China.\\
}

\history{Received on *****; revised on *****; accepted on *****}
\editor{Associate Editor: *****}
\begin{document}

\maketitle

\begin{abstract}
\section{Motivation}
Many computerized methods for RNA-RNA interaction structure
prediction have been developed. Recently, $O(N^6)$ time and $O(N^4)$
space dynamic programming algorithms have become available that
compute the partition function of RNA-RNA interaction complexes.
However, few of these methods incorporate the knowledge concerning
related sequences, thus relevant evolutionary information is often
neglected from the structure determination. Therefore, it is of
considerable practical interest to introduce a method taking into
consideration both thermodynamic stability and sequence covariation.
\section{Results}
We present the \emph{a priori} folding algorithm \texttt{ripalign},
whose input consists of two (given) multiple sequence
alignments (MSA). \texttt{ripalign} outputs (1) the partition function,
(2) base-pairing probabilities, (3) hybrid probabilities and (4) a set
of Boltzmann-sampled suboptimal structures consisting of canonical
joint structures that are compatible to the alignments.
Compared to the single sequence-pair folding algorithm \texttt{rip},
\texttt{ripalign} requires negligible additional memory resource.
Furthermore, we incorporate possible structure constraints as input
parameters into our algorithm.
\section{Availability}
The algorithm described here is implemented in C as part of the
\texttt{rip} package. The supplemental material, source code
and input/output files can freely be downloaded from
\url{http://www.combinatorics.cn/cbpc/ripalign.html}.
\section{Contact}
Christian Reidys \texttt{duck@santafe.edu}
\end{abstract}

{\bf Keywords }{multiple sequence alignment, RNA-RNA interaction,
joint structure, dynamic programming, partition function, base
pairing probability, hybrid, loop, RNA secondary structure.}
\maketitle
\section{Introduction}\label{S:Introduction}
RNA-RNA interactions play a major role at many different levels of
the cellular metabolism such as plasmid replication control, viral
encapsidation, or transcriptional and translational regulation. With
the discovery that a large number of transcripts in higher
eukaryotes are noncoding RNAs, RNA-RNA interactions in cellular
metabolism are gaining in prominence. Typical examples of
interactions involving two RNA molecules are snRNAs
\citep{Forne:96}; snoRNAs with their targets \citep{Bachellerie:02};
micro-RNAs from the RNAi pathway with their mRNA target
\citep{Ambros:04, Murchison:04}; sRNAs from \emph{Escherichia coli}
\citep{Hershberg:03, Repoila:03}; and sRNA loop-loop interactions
\citep{Brunel:02}. The common feature in many ncRNA classes,
especially prokaryotic small RNAs, is the formation of RNA-RNA
interaction structures that are much more complex than the simple
sense-antisense interactions.

As it is the case for the general RNA folding problem with
unrestricted pseudoknots \citep{Akutsu}, the RNA-RNA interaction
problem (RIP) is NP-complete in its most general form
\citep{Alkan:06,Mneimneh:07}. However, polynomial-time algorithms
can be derived by restricting the space of allowed configurations in
ways that are similar to pseudoknot folding algorithms
\citep{Rivas}. The simplest approach concatenates the two
interacting sequences and subsequently employs a slightly modified
standard secondary structure folding algorithm. The algorithms
\texttt{RNAcofold} \citep{Hofacker,Bernhart}, \texttt{pairfold}
\citep{Andronescu}, and \texttt{NUPACK} \citep{Ren} subscribe to
this strategy. A major shortcoming of this approach is that it
cannot predict important motifs such as kissing-hairpin loops. The
paradigm of concatenation has also been generalized to the
pseudoknot folding algorithm of \cite{Rivas}. The resulting model,
however, still does not generate all relevant interaction structures
\citep{Backofen}. An alternative line of thought is to neglect all
internal base-pairings in either strand and to compute the minimum
free energy (MFE) secondary structure for their hybridization under
this constraint. For instance, \texttt{RNAduplex} and
\texttt{RNAhybrid} \citep{rehmsmeier:04} follows this line of
thought. \texttt{RNAup} \citep{Mueckstein:05a,Mueckstein:08a} and
\texttt{intaRNA} \citep{Busch:08} restrict interactions to a single
interval that remains unpaired in the secondary structure for each
partner. These models have proved particularly useful for bacterial
sRNA/mRNA interactions \citep{Geissmann}.

\cite{Pervouchine:04} and \cite{Alkan:06} independently proposed MFE
folding algorithms for predicting the \emph{joint structure} of two
interacting RNA molecules with polynomial time complexity. In their
model, a ``joint structure'' means that the intramolecular
structures of each molecule are pseudoknot-free, the intermolecular
binding pairs are noncrossing and there exist no so-called
``zig-zags'', see supplement material (SM) for detailed definition.
The optimal joint structure is computed in $O(N^6)$ time and
$O(N^4)$ space via a dynamic programming (DP) routine.

A more reliable approach is to consider the partition function, which by
construction integrates over the Boltzmann-weighted probability space,
allowing for the derivation of thermodynamic quantities, like e.g.~equilibrium
concentration, melting temperature and base-pairing
probabilities. The partition function
of joint structures was independently derived by
\cite{Backofen} and \cite{rip:09} while the base-pairing probabilities
are due to \cite{rip:09}.

A key quantity here is the probability of hybrids, which
cannot be recovered from base pairing probabilities since the latter
can be highly correlated. \cite{rip2} presented a new hybrid-based
decomposition grammar,
facilitating the computation of the nontrivial hybrid-probabilities
as well as the Boltzmann sampling of RNA-RNA interaction structures.
The partition function of joint structures can be computed in
$O(N^6)$ time and $O(N^4)$ space and current implementations require
very large computational resources. \cite{Backofen:fast} recently
achieved a substantial speed-up making use of the observation that
the external interactions mostly occur between pairs of unpaired
regions of single structures. \cite{Chitsaz:09} introduced
tree-structured Markov Random Fields to approximate the joint
probability distribution of multiple $(\geq 3)$ contact regions.

Unfortunately, incompleteness of the underlying energy model, in
particular for hybrid- and kissing-loops, may result in prediction
inaccuracy. One way of improving this situation is to involve
phylogenetic information of multiple sequence alignments (MSA).

In an MSA homologous nucleotides are grouped in columns, where
homologous is interpreted in {\it both}: structural as well as
evolutionary sense. I.e.~a column of nucleotides occupies similar
structural positions and all diverge from a common ancestral
nucleotide. Also, many ncRNAs show clear signs of undergoing
compensatory mutations along evolutionary trajectories. In
conclusion, it seems reasonable to stipulate that a non-negligible
part of the existing RNA-RNA interactions contain preserved but
covarying patterns of the interactions \citep{Seemann:10}. Therefore
we can associate a consensus interaction structure to pairs of
interacting MSAs (see Section~\ref{S:basic}).

Along these lines \cite{Seemann:10} presented an algorithm
\texttt{PETcofold} for prediction of RNA-RNA interactions including
pseudoknots in given MSAs. Their algorithm is an extension of
\texttt{PETfold} \citep{Seemann:08} using elements of
\texttt{RNAcofold} \citep{Bernhart} and computational strategies for
hierarchical folding \citep{Gaspin:95, Jabbari:07}. However,
\texttt{PETcofold} is an approximation algorithm and further
differences between the two approaches will be discussed in
Section~\ref{S:discussion}.

Here, we present the algorithm \texttt{ripalign} which computes the
partition function, base-pairing as well as hybrid probabilities and
performs Boltzmann-sampling on the level of MSAs. \texttt{ripalign}
represents a generalization of \texttt{rip} to pairs of interacting
MSAs and a new grammar of canonical interaction structures. The
latter is of relevance since there are no isolated base pairs in
molecular complexes.

One important step consists in identifying the notion of a joint
structure compatible to a pair of interacting MSAs. Our notion is
based on the framework of \cite{Hofacker:02}, where a sophisticated
cost function capturing thermodynamic stability as well as sequence
covariation is employed. Furthermore \texttt{ripalign} is tailored
to take structure constraints, such as blocked nucleotides known
e.g.~from chemical probing, into account.

\begin{methods}
\section{Theory}
\subsection{Multiple sequence alignments and compatibility}\label{S:basic}

A MSA, $\bar{\mathbf{R}}$, consists of $m_{\bar{\mathbf{R}}}$ RNA
sequences of known species. Denoting the length of the aligned
sequences by $N$, $\bar{\mathbf{R}}$ constitutes a
$m_{\bar{\mathbf{R}}}\times N$ matrix, having $5'-3'$ oriented rows,
$\bar{\mathbf{R}}^{i}$ and columns, ${\bar{\mathbf{R}}}_{i}$. Its
$(i,j)$-th entry, $\bar{\mathbf{R}}^{i}_{j}$, is a nucleotide,
$\textbf{A},\textbf{U},\textbf{G},\textbf{C}$ or a gap denoted by
$\textbf{\texttt{.}}$.

For any pair $(\bar{\mathbf{R}},\bar{\mathbf{S}})$ we assume
that $\bar{\mathbf{S}}$ is a $m_{\bar{\mathbf{S}}}\times M$
matrix, whose rows carry $3'-5'$ orientation.

In the following we shall assume that a pair of RNA sequences can only
interact if they belong to the same species.
A pair $(\bar{\mathbf{R}},\bar{\mathbf{S}})$, can interact if for any
row $\bar{\mathbf{R}}^{i}$, there exist at least one row in
$\bar{\mathbf{S}}$ that can interact with $\bar{\mathbf{R}}^{i}$.

Given a pair of interacting MSAs
$(\bar{\mathbf{R}},\bar{\mathbf{S}})$, let $m$ be the total number
of potentially interacting pairs. \texttt{ripalign} exhibits a
pre-processing step which generates a $m\times N$-matrix
$\mathbf{R}$ and a $m\times M$-matrix $\mathbf{S}$ such that
$(\mathbf{R}^{i}, \mathbf{S}^{i})$ range over all $m$ potentially
interacting RNA-pairs, see Tab.~1 and the SM, Section~{1.2}.
\begin{table}[t]
\centering
\begin{tabular}{|l|c|l|c|l|c|c|}
  \hline
  sp.& {\ssm $\bar{\mathbf{R}}$} & sp.    &{\ssm $\bar{\mathbf{S}}$} &sp.&
  {\ssm$\mathbf{R}$} &{\ssm$\mathbf{S}$}\\ \hline
  $\theta_1$ & \textbf{\texttt{AGAACGGA}} & $\theta_1$ & \textbf{\texttt{GGGCCG}} & $\theta_1$ &\textbf{\texttt{AGAACGGA}}& \textbf{\texttt{GGGCCG}}\\
  $\theta_1$ & \textbf{\texttt{GAAACGGA}} & $\theta_1$ & \textbf{\texttt{AGUUAG}} & $\theta_1$ &\textbf{\texttt{AGAACGGA}}& \textbf{\texttt{AGUUAG}}\\
  $\theta_2$ & \textbf{\texttt{AGA.CGAC}} & $\theta_2$ & \textbf{\texttt{AGGCAG}} & $\theta_1$ &\textbf{\texttt{GAAACGGA}}& \textbf{\texttt{GGGCCG}}\\
             &                            & $\theta_2$ & \textbf{\texttt{..GUGG}} & $\theta_1$ &\textbf{\texttt{GAAACGGA}}& \textbf{\texttt{AGUUAG}}\\
             &                            &            &                          & $\theta_2$ &\textbf{\texttt{AGA.CGAC}}& \textbf{\texttt{AGGCAG}}\\
             &                            &            &                          & $\theta_2$ &\textbf{\texttt{AGA.CGAC}}& \textbf{\texttt{..GUGG}}\\
  \hline
\end{tabular}
\centerline{} \caption{\textbf{Preprocessing in \texttt{ripalign}:}
Given a pair of MSAs
 $(\bar{\mathbf{R}},\bar{\mathbf{S}})$, where
 $\bar{\mathbf{R}}$
  consists of three aligned RNA sequences of species (sp.) $\theta_1$
  or $\theta_2$.  $\bar{\mathbf{S}}$ in turn consists of four aligned
  sequences of species $\theta_1$ and $\theta_2$. Then we obtain
  the matrix-pair $({\mathbf{R}},{\mathbf{S}})$, where
  $(\mathbf{R}^{i},\mathbf{S}^{i})$, $1\leq i\leq 6$,
  ranges over all the six potentially interacting RNA-pairs.
  }
\end{table}
\label{T:1}

We shall refer in the following to $\mathbf{R}$ and $\mathbf{S}$ as
MSAs ignoring the fact that they have multiple sequences.

We proceed by defining joint structures that are compatible to a
fixed $({\mathbf{R}},{\mathbf{S}})$. To this end, let us briefly
review some concepts introduced in \cite{rip:09}.

A joint structure $J(R,S,I)$ is a graph consisting of\\
\textbf{(j1)} Two secondary structures $R$ and $S$, whose backbones
are drawn as horizontal lines on top of each other and whose arcs
are drawn in the upper and lower halfplane, respectively. We
consider $R$ over a $5'$ to $3'$ oriented backbone $(R_1,\dots,R_N)$
and $S$ over a $3'$ to $5'$ oriented backbone $(S_1,\dots,S_M)$ and
refer to any $R$- and $S$-arcs as interior arcs. \\
\textbf{(j2)}   An additional set $I$, of noncrossing arcs of the
form $R_iS_j$ (exterior arc), where $R_i$ and $S_j$ are unpaired in $R$ and $S$.\\
\textbf{(j3)} $J(R,S,I)$ contains no ``zig-zags'' (see SM).

The subgraph of a joint structure $J(R,S,I)$ induced by a pair of
subsequences $(R_i,R_{i+1},\dots,R_j)$ and $(S_h,
S_{h+1},\dots,S_\ell)$ is denoted by $J_{i,j;h,\ell}$. In
particular, $J(R,S,I)=J_{1,N;1,M}$ and $J_{i,j;h,\ell}\subset
J_{a,b;c,d}$ if and only if $J_{i,j;h,\ell}$ is a subgraph of
$J_{a,b;c,d}$ induced by $(R_i,\dots,R_j)$ and $(S_h,\dots,S_\ell)$.
In particular, we use $S[i,j]$ to denote the subgraph of
$J_{1,N;1,M}$ induced by $(S_i,S_{i+1}, \dots,S_j)$, where
$S[i,i]=S_{i}$ and $S[i,i-1]=\varnothing$.

Given a joint structure, $J_{a,b;c,d}$, a tight structure (TS),
$J_{i,j;h,\ell}$, \citep{rip:09} is a specific subgraph of
$J_{a,b;c,d}$ indexed by its type $\in
\{\circ,\bigtriangledown,\square,\bigtriangleup\}$, see
Fig.~\ref{F:typeins}. For instance, we use
$J^{\square}_{i,j;h,\ell}$ to denote a TS of type $\square$.
\begin{figure}
\begin{center}
  \includegraphics[width=\columnwidth]{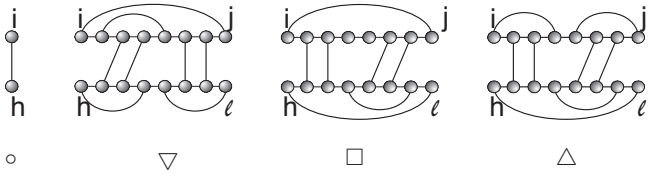}
\end{center}
\par\noindent
   $\circ$ \hspace*{0.2\columnwidth}
   $\bigtriangledown$\hspace*{0.25\columnwidth}
   $\square$ \hspace*{0.28\columnwidth}
   $\bigtriangleup$
\par\noindent
\caption{The four basic types of tight structures are given as
follows: $\circ:$ $\{R_iS_h\}=J_{i,j;h,\ell}$ and $i=j$, $h=\ell$;
$\bigtriangledown:$ $R_iR_j\in J_{i,j;h,\ell}$ and
      $S_{h}S_{\ell}\not\in J_{i,j;h,\ell}$;
$\square:$ $\{R_iR_j,S_{h}S_{\ell}\}\in J_{i,j;h,\ell}$;
$\bigtriangleup:$ $S_{h}S_{\ell} \in J_{i,j;h,\ell}$ and $R_iR_j\not
\in J_{i,j;h,\ell}$. } \label{F:typeins}
\end{figure}

A \emph{hybrid} is a joint structure
$J^{\mathsf{Hy}}_{i_1,i_\ell;j_1,j_\ell}$, i.e.~a maximal sequence
of intermolecular interior loops consisting of a set of exterior
arcs $(R_{i_1}S_{j_1}, \dots, R_{i_\ell}S_{j_\ell})$ where
$R_{i_h}S_{j_h}$ is nested within $R_{i_{h+1}}S_{j_{h+1}}$ and where
the internal segments $R[i_h+1,i_{h+1}-1]$ and $S[j_h+1,j_{h+1}-1]$
consist of single-stranded nucleotides only.  That is, a hybrid is
the maximal unbranched stem-loop formed by external arcs.

A joint structure $J(R,S,I)$ is called \textit{canonical} if and
only if: \\
{\bf (c1)} each stack in the secondary structures $R$ and $S$ is of
          size at least two, i.e.~there exist no isolated interior arcs,\\
{\bf (c2)} each hybrid contains at least two exterior arcs.\\
In the following, we always assume a joint structure to be canonical.

Next, we come to $(\mathbf{R},\mathbf{S})$-compatible joint structures.
In difference to single sequence compatibility, this notion
involves statistical information of the MSAs.

The key point consists in specifying under which conditions two
vertices contained in $(R_1,\dots,R_N, S_1,\dots,S_M)$ can pair.
This is obtained by a generalization of the \texttt{RNAalifold}
approach \citep{Hofacker:02}. We specify these conditions for
interior $(c_{i,j}^{\mathbf{R}})$, $(c_{i,j}^{\mathbf{S}})$ and
exterior pairs $(c_{i,j}^{\mathbf{R,S}})$ in
eq.~(\ref{E:c1})-(\ref{E:c3}). \\
For interior arcs $(R_i,R_j)$, let $\text{X,Y}\in\{\textbf{A},
\textbf{U},\textbf{G},\textbf{C}\}$. Let
$f_{ij}^{\mathbf{R}}(\text{XY})$ be the frequency of
$(\text{X},\text{Y})$ which exists in the $2$-column sub-matrix
$(\mathbf{R}_{i},\mathbf{R}_{j})$ as a row-vector and
\begin{equation}
C_{i,j}^{\mathbf{R}}=\sum_{\text{XY},\text{X}^{\prime}\text{Y}^{\prime}}
f_{ij}^{\mathbf{R}}(\text{XY})D^{\mathbf{R}}_{\text{XY},\text{X}^{\prime}
\text{Y}^{\prime}}f^{\mathbf{R}}_{ij}(\text{X}^{\prime}\text{Y}^{\prime}).
\end{equation}
Here XY and X$'$Y$'$ independently range over all 16 elements
of $\{\textbf{A},\textbf{U},\textbf{G},\textbf{C}\}\times\{\textbf{A},
\textbf{U},\textbf{G},\textbf{C}\}$ and
$D^{\mathbf{R}}_{\text{XY},\text{X}'\text{Y}'}=d_{H}(\text{XY},
\text{X}'\text{Y}')$, i.e.~the Hamming distance between XY and
X$'$Y$'$ in case of XY and X$'$Y$'$ being Watson-Crick, or
$\textbf{G}\textbf{U}$ wobble base pair and 0, otherwise.
Furthermore, we introduce $q_{i,j}^{\mathbf{R}}$ to deal with the
inconsistent sequences
\begin{equation}
q_{i,j}^{\mathbf{R}}=1-\frac{1}{m}\sum_{h}\{\Pi_{i,j}^{h}(\mathbf{R})+
\delta(\mathbf{R}_{i}^{h},\text{gap})
\delta(\mathbf{R}_{j}^{h},\text{gap})\},
\end{equation}
where $\delta(x,y)$ is the Kronecker delta and
$\Pi_{i,j}^{h}(\mathbf{R})$ is equal to 1 if $\mathbf{R}^{h}_{i}$
and $\mathbf{R}^{h}_{j}$ are Watson-Crick or $\textbf{G}\textbf{U}$
wobble base pair and 0, otherwise.
Now we obtain $B_{i,j}^{\mathbf{R}}=C_{i,j}^{\mathbf{R}}-
\phi_{1}q_{i,j}^{\mathbf{R}}$. Based on sequence data, the threshold
for pairing $B^{\mathbf{R}}_{*}$ as well as the weight of inconsistent
sequences $\phi_{1}$ are computed we have
\begin{equation}\label{E:c1}
(c_{i,j}^{\mathbf{R}})\quad B^{\mathbf{R}}_{i,j}\geq
B^{\mathbf{R}}_{*}
\end{equation}
The case of two positions $S_{i}$ and $S_{j}$ is completely analogous
\begin{equation}\label{E:c2}
(c_{i,j}^{\mathbf{S}})\quad B^{\mathbf{S}}_{i,j}\geq
B^{\mathbf{S}}_{*},
\end{equation}
where $B^{\mathbf{S}}_{i,j}$ and $B^{\mathbf{S}}_{*}$ are
analogously defined.

As for $(c_{i,j}^{\mathbf{R},\mathbf{S}})$ a further observation
factors in: since many ncRNA show clear signs of undergoing
compensatory mutations in the course of evolution \citep{Seemann:10,Marz:08},
we postulate the existence of a non-negligible amount of RNA-RNA
interactions containing conserved pairs, consistent mutations,
compensatory mutations as well as inconsistent mutations.  Based on
this observation we arrive at
\begin{equation}\label{E:c3}
(c_{i,j}^{\mathbf{R},\mathbf{S}})\quad
B^{\mathbf{R},\mathbf{S}}_{i,j}\geq B^{\mathbf{R},\mathbf{S}}_{*},
\end{equation}
where $B^{\mathbf{R},\mathbf{S}}_{i,j}$ and
$B^{\mathbf{R},\mathbf{S}}_{*}$ are analogously defined as the case
for $B^{\mathbf{R}}_{i,j}$ and $B^{\mathbf{R}}_{*}$.

A joint structure $J$ is compatible to $(\mathbf{R},\mathbf{S})$ if
for any $J$-arc, the corresponding intra- or inter-positions can
according to eq.~(\ref{E:c1})-(\ref{E:c3}) pair.
\subsection{Energy model}\label{S:loop}
According to \cite{rip:09} joint structures can be decomposed into
disjoint loops. These loop-types include standard hairpin-, bulge-,
interior- and multi-loops found in RNA secondary structures as well
as \emph{hybrid} and \emph{kissing-loops}. Following the energy
parameter rules of \cite{Mathews}, the energy of each loop can be
obtained as a sum of the energies associated with non-terminal
symbols, i.e.~graph properties (sequence independent) and an
additional contributions which depend uniquely on the terminal bases
(sequence dependent).
\begin{figure}[t]
\begin{center}
  \includegraphics[width=0.4\columnwidth]{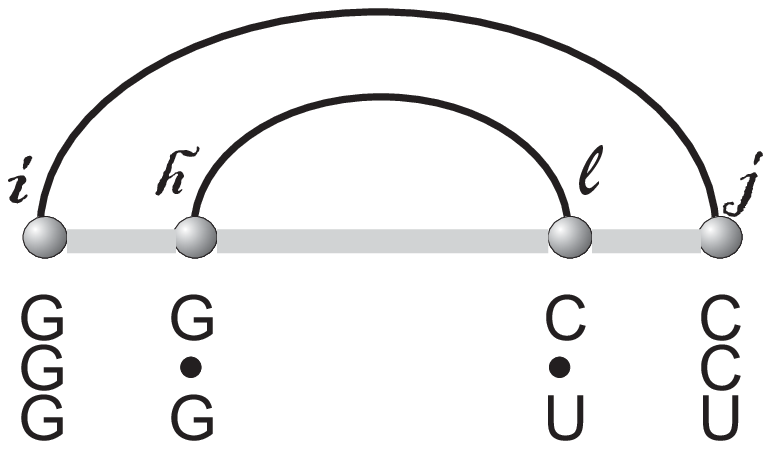}
\end{center}
\caption{\textbf{Interior loop energy:} An interior loop formed by
$R_{i}R_{j}$ and $R_{h}R_{\ell}$, where $i<h<\ell<j$ are the
alignment positions. Grey bands are used to denote the positions we
omit between segment $(i,h)$, $(h,\ell)$ and $(\ell,j)$.}
\label{F:loop}
\end{figure}

Suppose we are given a joint structure $J$, compatible to a pair
$\mathcal{P}=(\mathbf{R},\mathbf{S})$. Let $L\in J$ be a loop and
let $\mathcal{F}_{L,i}$ represent the loop energy of the $i$-th
interaction-pair $(\mathbf{R}^{i},\mathbf{S}^{i})$. Then the loop
energy of $\mathcal{P}$ is
\begin{equation}\label{E:loopenergy}
\mathcal{F}_{L,\mathcal{P}} = 1/m \sum_{i}\mathcal{F}_{L,i}.
\end{equation}
We consider the energy of the structure as the sum of all loop
contributions:
\begin{equation}\label{E:energy1}
\mathcal{F}_{J}=\sum_{L\in J}\mathcal{F}_{L,\mathcal{P}}.
\end{equation}
To save computational resources, gaps are treated as bases in
\texttt{ripalign}. Thus only alignment positions contribute as
indices and loop sizes. Since no measured energy parameters for
nonstandard base-pairs are available at present time, additional
terminal-dependent contributions for the latter are ignored. For
instance, let ${\sf Int}_{i,j;h,l}$ denote an interior loop formed
by $R_{i}R_{j}$ and $R_{h}R_{\ell}$ and
$\mathcal{F}_{\textsf{Int},\mathcal{P}}^{i,j;h,\ell}$ denote the
free energy of $\textsf{Int}_{i,j;h,l}$ with respect to the aligned
sequences in $\mathcal{P}$. Then $\mathcal{F}_{\textsf{Int},
\mathcal{P}}^{i,j;h,\ell}$ associated to the three aligned
subsequences of Fig.~\ref{F:loop} reads
\begin{equation}\label{E:loopexample}
\mathcal{F}^{\textsf{Int},\mathcal{P}}_{i,j;h,\ell}=\frac{1}{3}(3G^{\sf
Int}_{i,j;h,\ell}+G_{*,\textbf{G,C;G,C}}^{\sf
Int}+G_{*,\textbf{G,U;G,U}}^{\sf
Int}+G_{*,\textbf{G,C;gap,gap}}^{\sf Int}).
\end{equation}
Here $G^{\sf Int}_{i,j;h,\ell}$ represents contributions related exclusively
to the positions of the interior loop while $G^{\sf Int}_{*,\textbf{A,B;C,D}}$
represents additional contributions related to the specific nucleotides which
form the interior loop. We set $G_{*,\textbf{G,C;gap,gap}}^{\sf Int}$ to be
zero.
\begin{figure}
\begin{center}
  \includegraphics[width=\columnwidth]{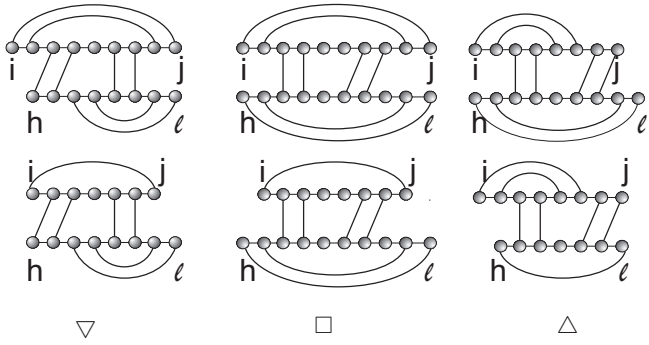}
\end{center}
\begin{center}
\par\noindent
   $\bigtriangledown$\hspace*{0.33\columnwidth}
   $\square$ \hspace*{0.33\columnwidth}
   $\bigtriangleup$
\par\noindent
\end{center}
\caption{\textbf{Examples of two TS-types.} We display
$\bigtriangledown$,
  $\square$, or $\bigtriangleup$-tight structures: Type cc (top) and
  Type c (bottom). } \label{F:typeins2}
\end{figure}
\begin{figure}[t]
\begin{center}
\includegraphics[width=1\columnwidth]{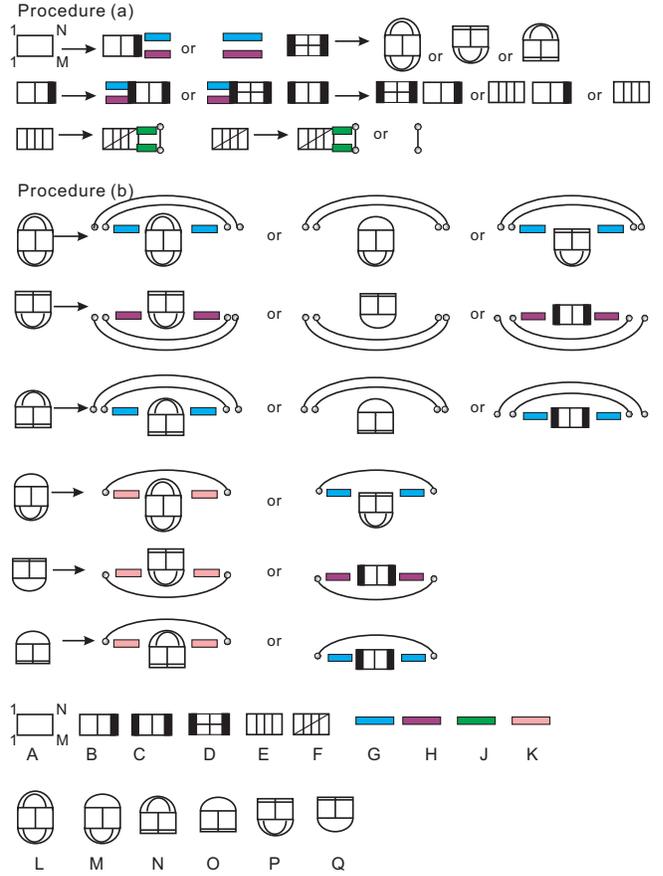}
\end{center}
\caption{\textbf{Grammar:} Illustration of the decomposition of
$J_{1,N;1,M}$, DTS, RTS and hybrids in Procedure (a) and of tight
structures in Procedure (b). In the bottom row the symbols for the
16 distinct types of structural components are listed:
  \textbf{A}: arbitrary joint structure $J_{1,N;1,M}$ (canonical);
  \textbf{B}: right-tight structures $J^{RT}_{i,j;r,s}$;
  \textbf{C}: double-tight structure $J^{DT}_{i,j;r,s}$;
  \textbf{D}: tight structure $J^{\bigtriangledown, cc}_{i,j;h,\ell}$,
    $J^{\bigtriangleup, cc}_{i,j;h,\ell}$ or $J^{\square, cc}_{i,j;h,\ell}$;
  \textbf{E}: hybrid structure $J^{\sf Hy}_{i,j;h,\ell}$;
  \textbf{F}: substructure of a hybrid $J^{\sf
  h}_{i,j;h,\ell}$ such that $R_{i}S_{j}$ and $R_{h}S_{\ell}$ are
  exterior arcs and $J^{\sf
  h}_{i,j;h,\ell}$ itself is not a hybrid since it is not maximal;
  \textbf{G}, \textbf{H}: maximal secondary structure
     segments $R[i,j]$, $S[r,s]$;
  \textbf{J}: isolated segment $R[i,j]$ or $S[h,\ell]$;
  \textbf{K}: maximal secondary structure
     segments appear in pairs such that at
     least one of them is not empty.
  \textbf{L}: tight structure $J^{\square,cc}_{i,j;r,s}$;
  \textbf{M}: tight structure $J^{\square,c}_{i,j;r,s}$;
  \textbf{N}: tight structure $J^{\bigtriangledown,cc}_{i,j;r,s}$;
  \textbf{O}: tight structure $J^{\bigtriangledown,c}_{i,j;r,s}$;
  \textbf{P}: tight structure $J^{\bigtriangleup,cc}_{i,j;r,s}$;
  \textbf{Q}: tight structure $J^{\bigtriangleup,c}_{i,j;r,s}$.
   } \label{F:grammar}
\end{figure}
\subsection{The grammar of canonical joint structures and the
            partition function}\label{S:grammar}
The partition function algorithm is easily extended to work with the
modified energy functions given in eq.~(\ref{E:energy1}).  The
reformulation of the original hybrid-grammar into a grammar of
canonical joint structures represents already for single interaction
pairs a significant improvement in prediction quality. The original
\texttt{rip}-grammar would oftentimes encounter joint structures
having a hybrid composed by a single isolated exterior arc, see
Fig.~\ref{F:comversion}.\\
In order to decompose canonical joint structures via the unambiguous
grammar introduced in Section~\ref{S:grammar}, we distinguish the
two types (Type cc and Type c) of TS's of type $\bigtriangledown$,
$\bigtriangleup$ or $\square$. Given a TS of type
$\bigtriangledown$, denoted by $J^{\bigtriangledown}_{i,j;h,\ell}$,
we write depending on whether $R_{i+1}R_{j-1}\in
J^{\bigtriangledown}_{i,j;h,\ell}$,
$J^{\bigtriangledown,cc}_{i,j;h,\ell}$ and $J^{\bigtriangledown,
c}_{i,j;h,\ell}$, respectively. Analogously, we define $J^{\square,
cc}_{i,j;h,\ell}$, $J^{\square,c}_{i,j;h,\ell}$ and
$J^{\bigtriangleup,cc}_{i,j;h,\ell}$,
$J^{\bigtriangleup,c}_{i,j;h,\ell}$, see Fig.~\ref{F:typeins2}.\\
Fig.~\ref{F:grammar} summarizes the two basic steps of the
canonical-grammar: (I) {\emph interior arc-removal} to reduce TS,
and (II) {\emph block-decomposition} to split a joint structure into
two smaller blocks. The key feature here is, that since $J$ is
canonical, the smaller blocks are still canonical after
block-decomposition. Each decomposition step displayed in
Fig.~\ref{F:grammar} results in substructures which eventually break
down into generalized loops whose energies can be directly computed.
More details of the decomposition procedures are described in
Section~2 of the SM, where we prove that for any canonical joint
structure $J$, there exists a unique decomposition-tree
(parse-tree), denoted by $T_{J}$, see Fig.~\ref{F:tree}.
\begin{figure}[t]
\begin{center}
  \includegraphics[width=\columnwidth]{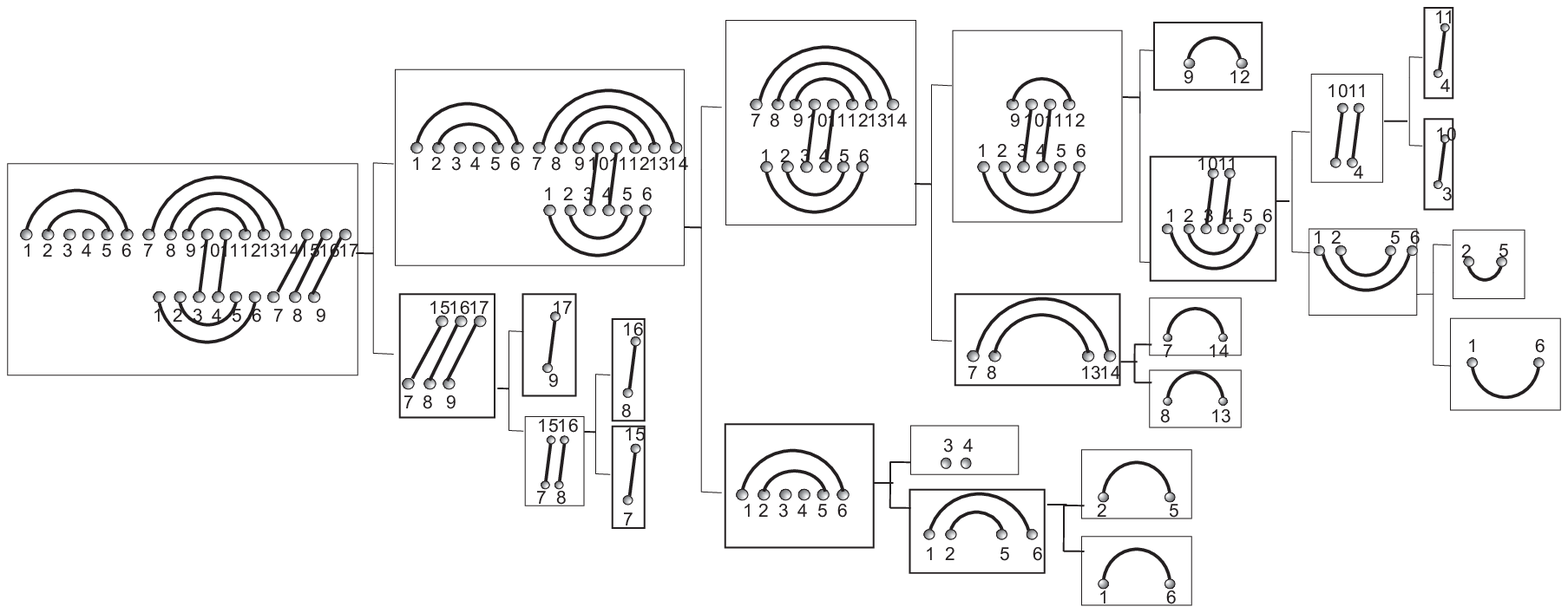}
\end{center}
\caption{\textbf{Example of the parse tree.} The parse tree of the
canonical joint structure $J_{1,17;1,9}$.} \label{F:tree}
\end{figure}
\subsection{Probabilities and the Boltzmann
Sampling}\label{S:prosam}
A dynamic programming scheme for the computation of a partition
function implies a corresponding computation of probabilities of
specific substructures is obtained ``from the outside to the inside"
and a stochastic backtracing procedure that can be used to sample
from the associated distribution \citep{McCaskill, Ding:03, rip2}.
We remark that the time complexity does not increase linearly as a
function of $m$ (see SM Table.~5).\\
Along the lines of the design of the Vienna software package
\citep{Hofacker}, \texttt{ripalign} now offers the following
features as optional input parameters:\\
{\sf (1)} a position $i$ can be restricted to form an interior or an
          exterior arc. (denoted by ``$-$'' and ``\,\textasciicircum\,'',
          respectively);\\
{\sf (2)} a position $i$ can be forced to be unpaired (denoted by ``x''); \\
{\sf (3)} a position $i$ can be restricted to form an (interior or
          an exterior) arc with {\it some} position $j$
          (denoted by ``$*$''); \\
{\sf (4)} a pair of positions $i$ and $j$ can be forced to form an
          interior or exterior arc (denoted by ``$(\,)$'' or ``$[\,]$'',
          respectively).\\
However, the above features are optional. Thus  \texttt{ripalign}
can deal with both scenarios: the absence of any \emph{a priori}
information and the existence of specific information, e.g~the
location of the Sm-binding site, see Fig.~\ref{F:comversion}.
\section{Results and discussion}\label{S:conclusion}
In this paper we present an \emph{a priori} $O(N^6)$ time and
$O(N^4)$ space dynamic programming algorithm \texttt{ripalign},
whose input consists of a pair of interacting MSAs.
\texttt{ripalign} requires only marginally more computational
resources but is, without doubt, still computationally costly.
Approximation algorithms are much faster, for instance
\texttt{PETcofold} \citep{Seemann:10}, having a time complexity of
$O(m\,(N+M)^3\,n)$, where $m$ is the number of sequences in MSA, $N$
and $M$ being the sequence lengths of the longer and shorter
alignment, respectively, and $n<N/2$ is the number of iterations for
the adaption of the threshold value to find likely partial secondary
structures. Their basic assumption is that the two secondary
structures fold independently and that intra-loop evaluation
differences are negligible.  The flip-side of reducing the
complexity of a folding problem by introducing additional
assumptions, is however, the uncertainty of the quality of the
solution. Point in case here is that the two secondary structures
did not evolve independently, but rather correlated by means of
their functional interaction. We remark that \texttt{ripalign}
(within its complexity limitations) is capable to describe the space
of RNA interaction structures, for instance via Boltzmann sampling,
in detail and transparency.\\
\texttt{ripalign} represents significant improvements in the
following
aspects:\\
{\bf (a)} we incorporate evolutionary factors into the
          RNA-RNA interaction structure prediction via alignments as input,\\
{\bf (b)} we introduce the grammar of canonical joint
          structures of interacting-alignments,\\
{\bf (c)} we \emph{a priori} factor in structural-constraints, like
          for instance, knowledge on Sm-binding sites.\\
Below we shall discuss {\bf (a), (b)} and {\bf (c)} in more detail in the
context of concrete examples. All the MSAs involving in {\bf
(a), (b)} and {\bf (c)} are listed in SM, Section 2.

{\bf (a): The \emph{fhlA}/\emph{OxyS} interaction}\\
The \emph{OxyS} RNA represses \emph{fhlA} mRNA translation
initiation through base-pairing with two short
sequences\cite{Argaman:00}, one of which overlaps the ribosome
binding sequence and the other resides further downstream, within
the coding region of \emph{fhlA}. Our algorithm predicts correctly
both interaction sites based on MSAs, see Fig.~\ref{F:singlevsmsa}.
In addition, most predicted stacks in the secondary structures of
\emph{fhlA} and \emph{OxyS} agree well with the most frequent
Bolztmann sampled structure. Two more hybrids, $J^{\sf
Hy}_{56,59;41,44}$ and $J^{\sf Hy}_{81,83;48,50}$ are predicted in
our output. The two additional contact regions, identified in the
partition function, exhibit a significantly lower probability. An
additional hairpin over $R[72,89]$ is predicted in \emph{fhlA},
instead of the unpaired segment occurring in the natural structure,
can be understood in the context of minimizing free energy.
Comparing the prediction based on the MSAs
(Fig.~\ref{F:singlevsmsa}, middle) with the one based on the
consensus sequence
(Fig.~\ref{F:singlevsmsa}, bottom), we observe: \\
(1) the secondary structure of \emph{fhlA} agrees better with the
annotation joint structure (Fig.~\ref{F:singlevsmsa}, top), \\
(2) the leftmost hybrid agrees better with that of the annotated
structure. \\
(3) the binding-site probability (see SM, Section~5, eq.~(5.5))
of the leftmost hybrid increases by nearly 40\%. \\
On the flip side, due to the gaps in seven out of eight subsequences
induced by $R[98,102]$ (Column 98-102 in \emph{fhlA}), the
prediction quality of the right-most hybrid and its corresponding
contact-region probability decreases slightly.\\
Let us next contrast our results with those of \texttt{PETcofold},
see Fig.~\ref{F:comfhlA}. The latter predicts \emph{one} of the two
interaction sites. The second site is predicted subject to the
condition that constrained stems were not extended
\citep{Seemann:10}. It can furthermore be observed that in order to
predict the second hybrid, at the same time the secondary structures
prediction of both \emph{fhlA} and \emph{OxyS} gets worse.
\texttt{ripalign} predicts both: the interaction sites situated in
\emph{fhlA} and comes close to predicting the secondary structures
of \emph{fhlA} as well as \emph{OxyS} without any additional
constraints.

{\bf (b): The \emph{SmY-10}/\emph{SL-1} interaction of \emph{C.
elegans}}\\
\cite{MacMorris:07} stipulated that \emph{SmY-10} RNA, possible
involved in \emph{trans-}splicing, interacts with the splice leader
RNA (\emph{SL1} RNA). In Fig.~\ref{F:comversion}, we show that the
Sm-binding sites (colored in red) of the RNA molecules \emph{SmY-10}
and \emph{SL-1} are $R[56,62]$ and $S[25,31]$, respectively. In
Fig.~\ref{F:comversion}, the top structure is being predicted by
\texttt{rip} \citep{rip2}. We observe that firstly a stack in
\emph{SmY-10} consisting of the single arc $R_{24}S_{67}$ and
secondly the nucleotides of the Sm-binding sites form intra base
pairs. The canonical grammar presented here restricts the
configuration ensemble to canonical joint structures, resulting in
the structure presented in Fig.~\ref{F:comversion} (middle) in which
the peculiar isolated interaction arc disappears. However, the
nucleotides of the Sm-binding sites still form either intra or
inter-molecular base pairs. Incorporating the structural constraints
option we derive the bottom structure displayed in
Fig.~\ref{F:comversion}. Here the Sm-binding sites are
single-stranded. In Table.~\ref{T:2} we elaborate this point further
and show that the combination of canonical grammar and structural
constraints eliminate unwanted hybrids and ``free'' the nucleotides
attributed to Sm-binding sites of unwanted interactions.
\begin{figure}
\begin{center}
\cite{Argaman:00}
\includegraphics[angle=90,width=1\columnwidth]{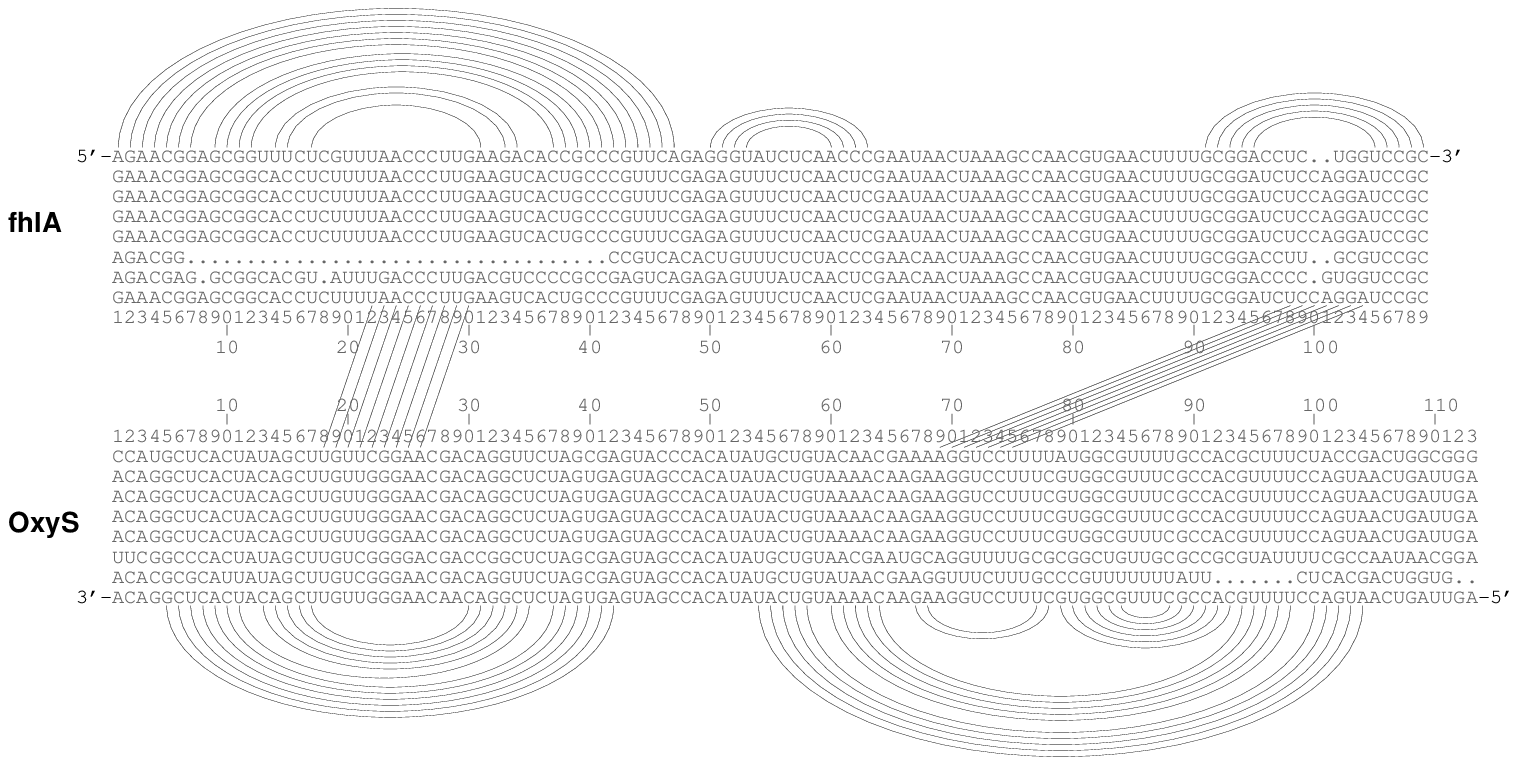}
\texttt{ripalign}: MSA-input
\includegraphics[angle=90,width=1\columnwidth]{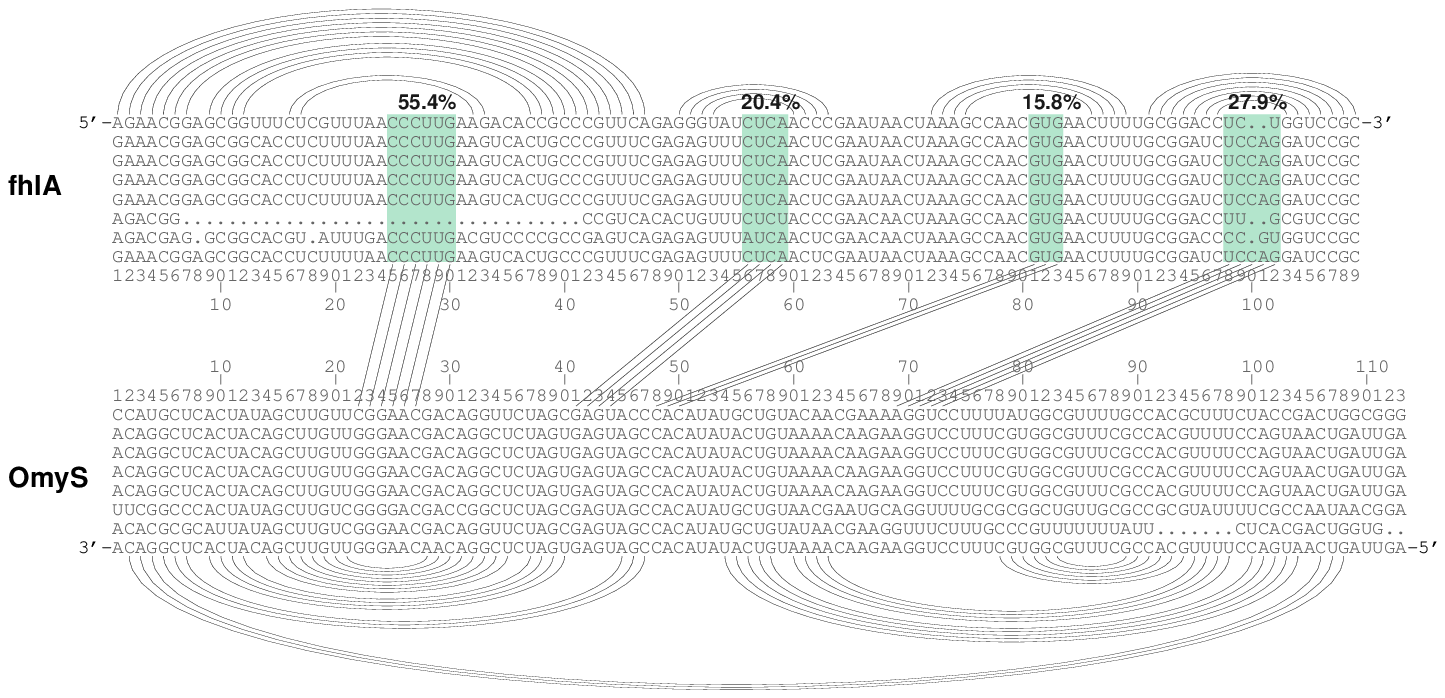}
\texttt{ripalign}: Single-sequences input
\includegraphics[angle=90,width=1\columnwidth]{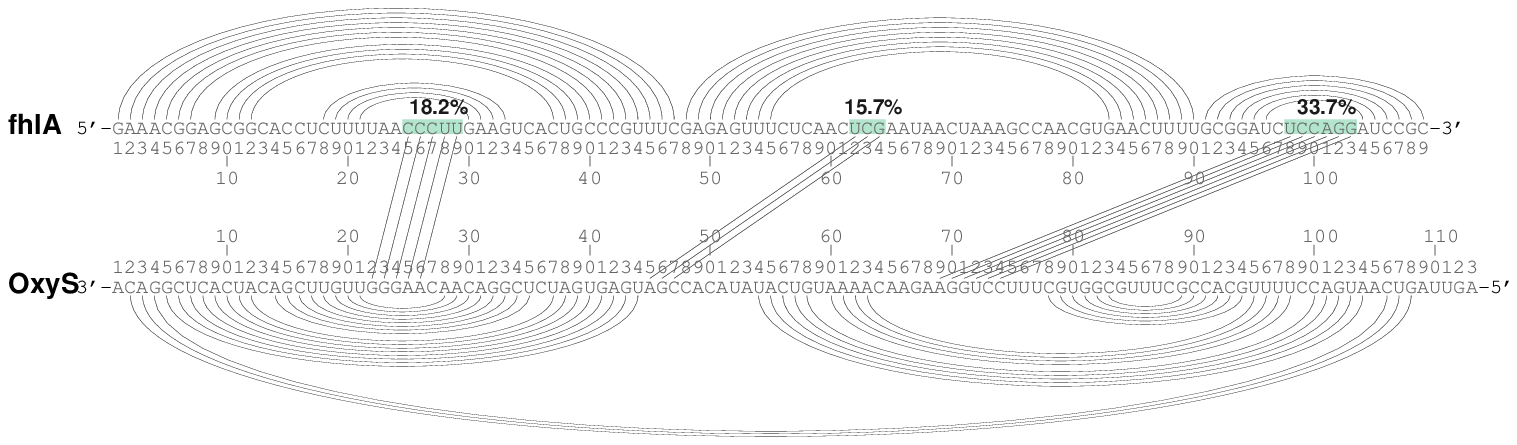}
\end{center}
\caption{\textbf{Improvement of \texttt{prediction} via
incorporating evolutionary history.} Top: the annotated structure of
the \emph{fhlA}/\emph{OxyS} interaction \cite{Argaman:00}; Middle:
the joint structure predicted by \texttt{ripalign} with MSAs as
input; Bottom: the joint structure predicted by \texttt{ripalign}
with the consensus sequences of MSAs as input. The target site
(green boxes) probabilities (defined in SM Section.~5, eq.~(5.5))
computed by \texttt{ripalign} are annotated explicitly if $>10\%$ or
just by $\leq 10\%$, otherwise. For instance, the probability of the
left-most contact region $R[25,30]$ in \emph{fhlA} (middle) is
$55.4\%$.} \label{F:singlevsmsa}
\end{figure}
\begin{figure}[t]
\begin{center}
\texttt{PETcofold} without the extension of the constrained stems
\includegraphics[angle=90,width=1\columnwidth]{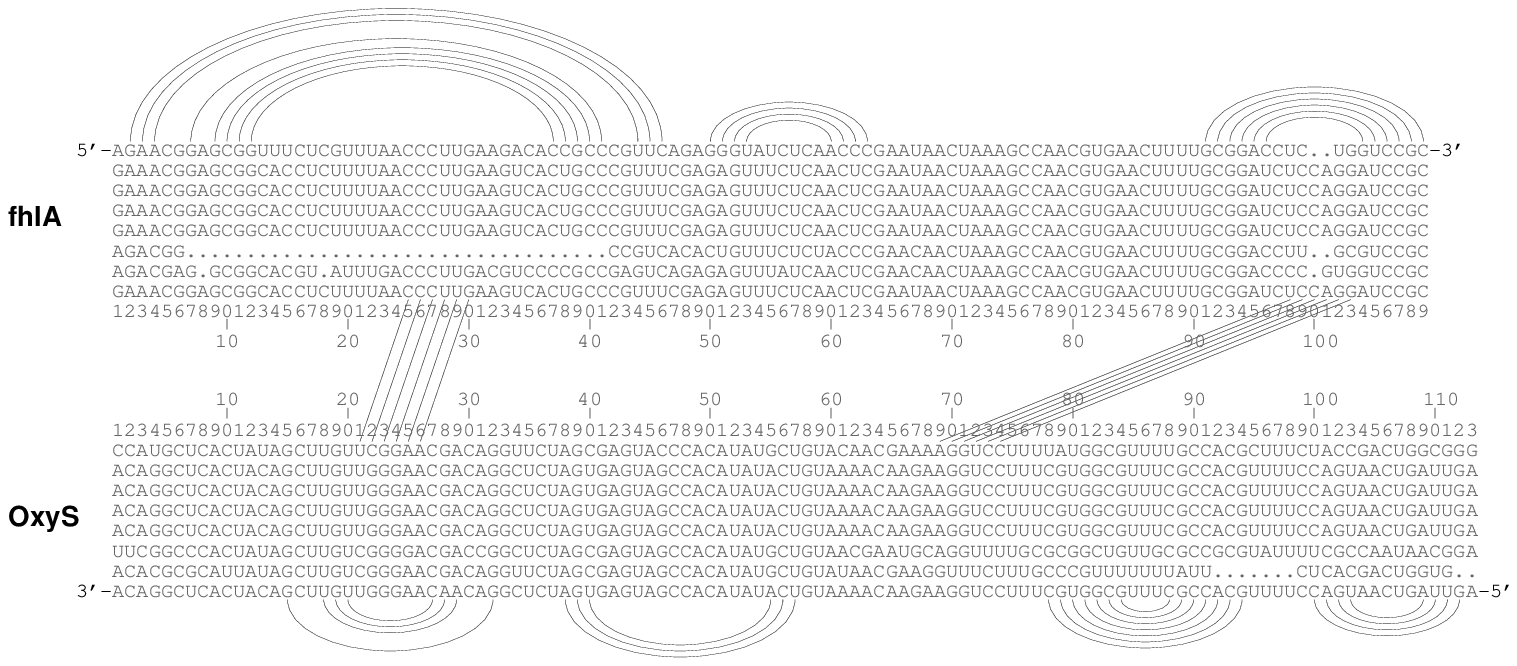}
\texttt{PETcofold} with the extension of the constrained stems
\includegraphics[angle=90,width=1\columnwidth]{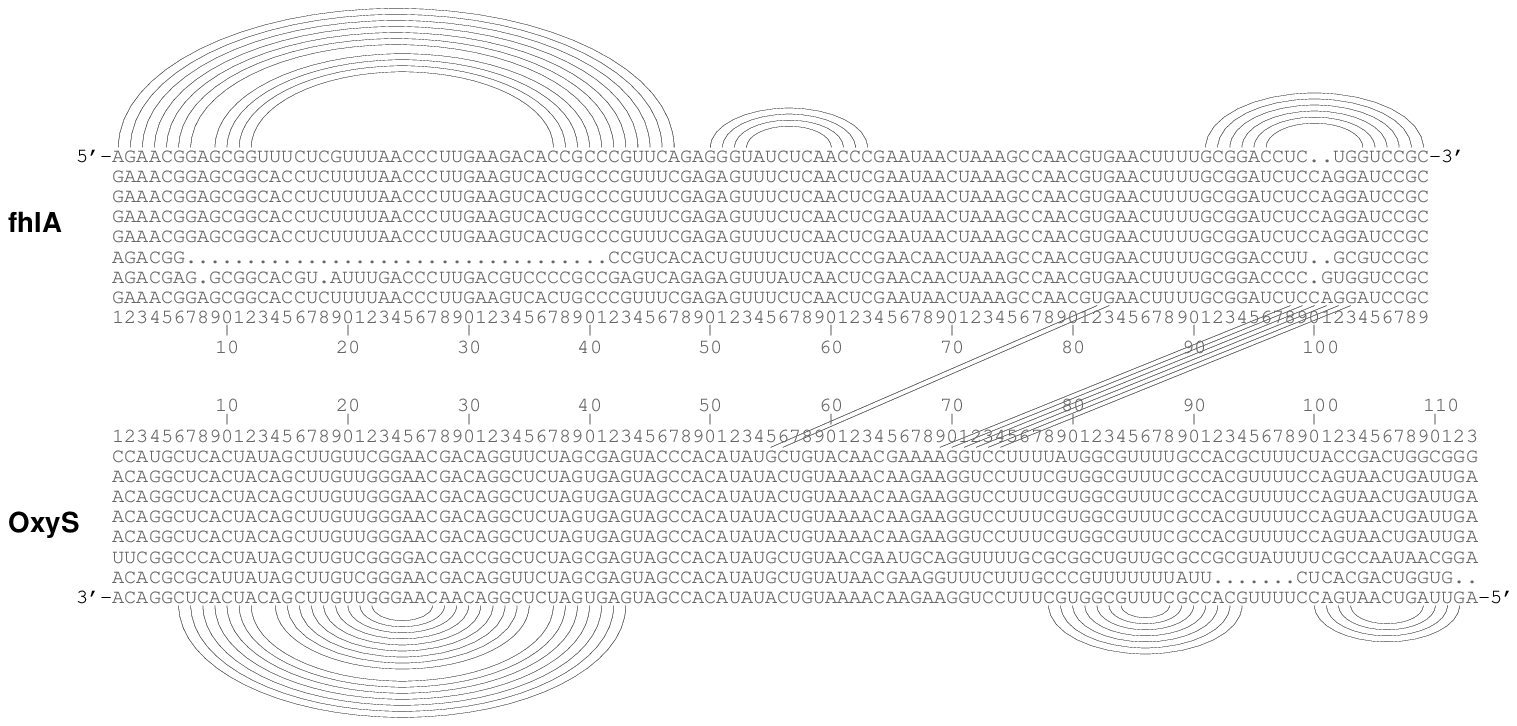}
\end{center}
\caption{\textbf{The joint structures of \emph{fhlA}/\emph{OxyS}
predicted by \texttt{PETcofold}:} The prediction was performed (top)
without and (bottom) with the extension of the constrained stems
based on the same MSAs showed in Fig.~\ref{F:singlevsmsa}. Here, the
extension of constrained stems is a specific programming-technique
of \cite{Seemann:10} to avoid incomplete stems appear in their
prediction result.} \label{F:comfhlA}
\end{figure}
\begin{figure}[t]
\begin{center}
\texttt{rip}
\includegraphics[angle=90,width=1\columnwidth]{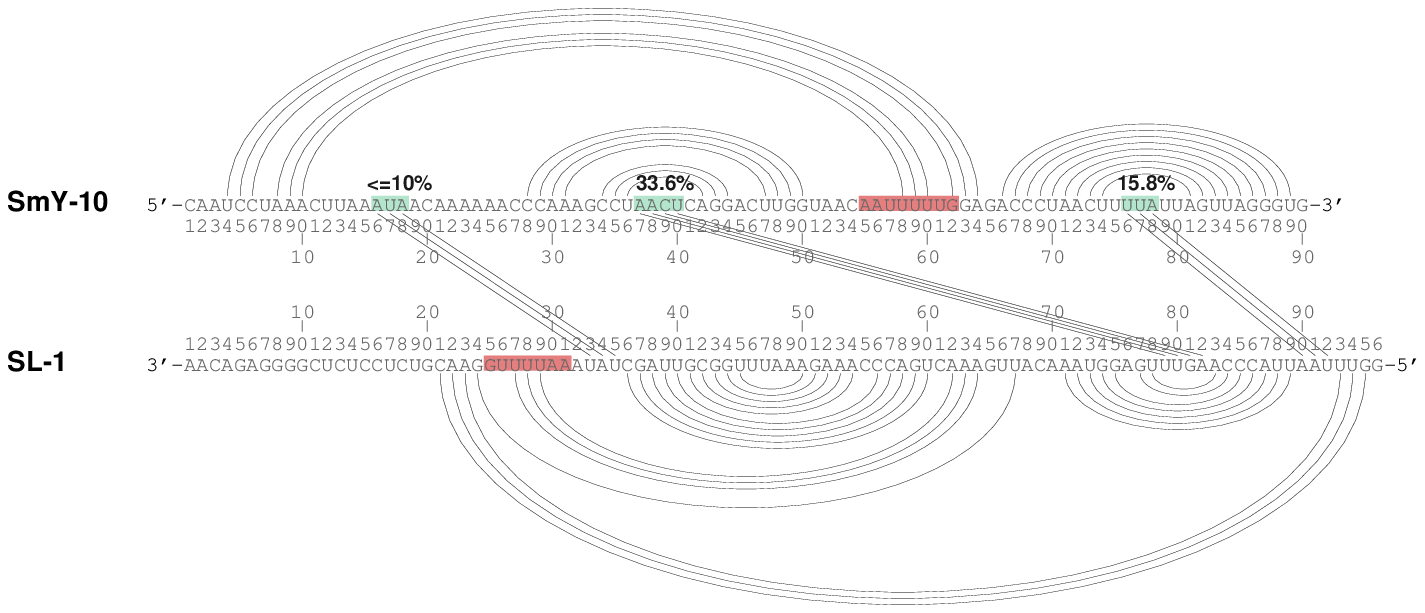}
\texttt{ripalign} without structure-constraint
\includegraphics[angle=90,width=1\columnwidth]{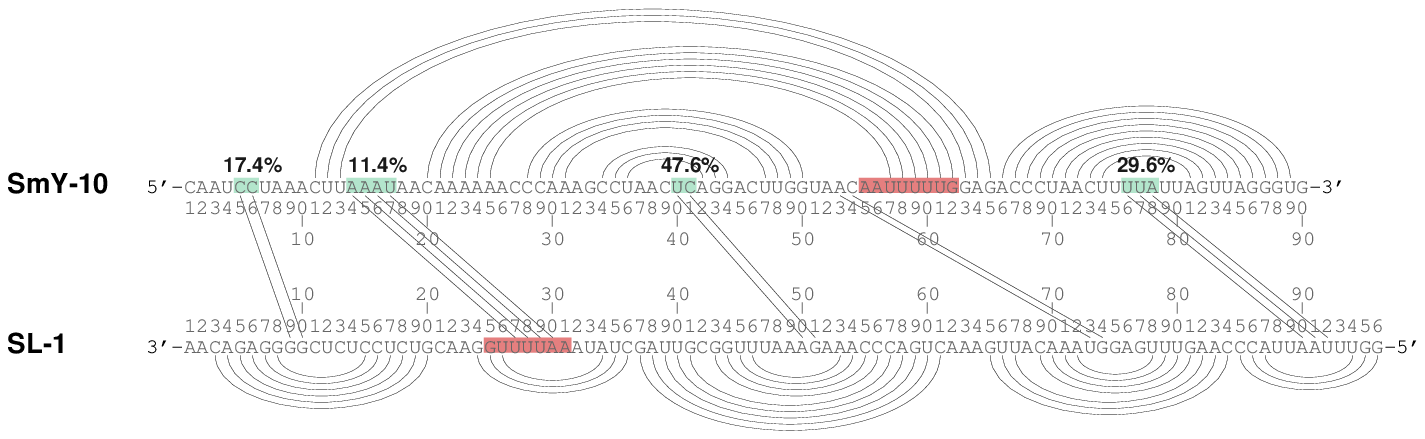}
\texttt{ripalign} with structure-constraint
\includegraphics[angle=90,width=1\columnwidth]{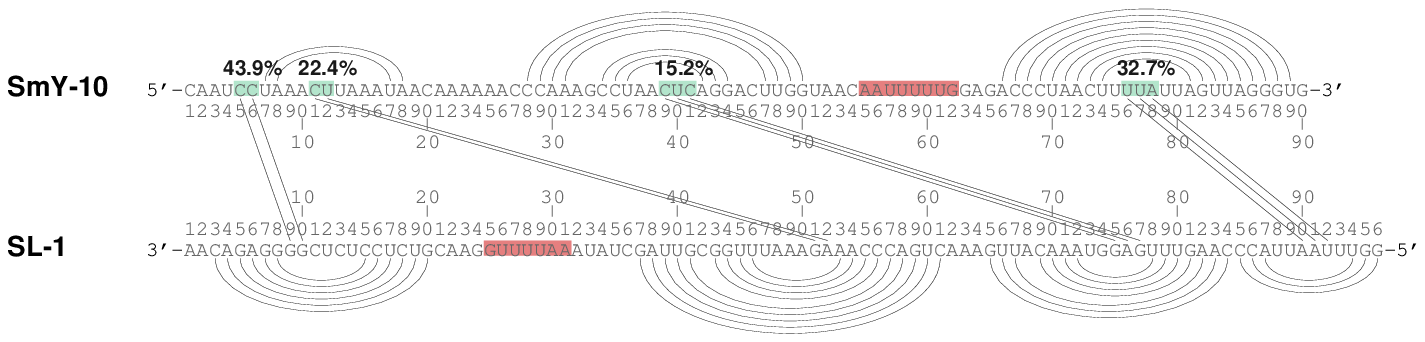}
\end{center}
\caption{\textbf{\texttt{ripalign} versus \texttt{rip}:} Interaction
  of two specific RNA molecules, \emph{SL1} and \emph{SmY-10} of
  \emph{Caenorhabditis elegans}. The Sm-binding sites (colored in red)
  in the
  RNA molecules \emph{SmY-10} and \emph{SL-1} are
  $\textbf{5'-AAUUUUUG-3'}(R[56,62])$ and
  $\textbf{3'-GUUUUAA-5'}(S[25,31])$, respectively. The joint structure
  contain a single interior arc $R_{24}S_{67}$(top) is predicted by
  \texttt{rip} implemented by \cite{rip2}. The
  joint structure (middle) is predicted by \texttt{ripalign} without any
  structural constraint. The joint structure (bottom) is predicted by
  \texttt{ripalign} under the structural constraints that
  $\textbf{5'-AAUUUUUG-3'}(R[56,62])$ and
  $\textbf{3'-GUUUUAA-5'}(S[25,31])$ are Sm-binding sites in the RNA
  molecules \emph{SmY-10} and \emph{SL-1}, respectively. The target site
  (green boxes) probabilities computed by \texttt{ripalign} are
  annotated explicitly if $>10\%$ or just by $\leq 10\%$,
  otherwise.}
\label{F:comversion}
\end{figure}
\begin{table}[t]
\centering
\begin{tabular}{|l|l|l|l|}
  \hline
  & I&II&III\\ \hline
   1&$J^{\sf {Hy}}_{37,40;79,82}$&$J^{\sf {Hy}}_{40,41;50,51}$&
   $J^{\sf {Hy}}_{5,6;9,10}$\\
  \hline
   2&$J^{\sf {Hy}}_{40,41;50,51}$&$J^{\sf {Hy}}_{39,40;51,52}$
  &$J^{\sf {Hy}}_{76,78;90,92}$\\
  \hline
   3&$J^{\sf {Hy}}_{76,78;90,92}$&$J^{\sf {Hy}}_{76,78;90,92}$
  &$J^{\sf {Hy}}_{37,40;79,82}$\\
  \hline
  4&$\mathbf{R_{11}S_{10}}$&$J^{\sf {Hy}}_{11,12;9,10}$&
  $J^{\sf {Hy}}_{78,80;89,91}$\\
  \hline
  5&$J^{\sf {Hy}}_{16,18;33,35}$&$J^{\sf {Hy}}_{78,80;89,91}$&
  $J^{\sf {Hy}}_{11,12;51,52}$\\
  \hline
  6&$J^{\sf {Hy}}_{54,57;65,68}$&$J^{\sf {Hy}}_{54,57;65,68}$&
  $J^{\sf {Hy}}_{16,17;47,48}$\\
  \hline
\end{tabular}
\caption{\textbf{Top 6 probable hybrids predicted by \texttt{rip}
and \texttt{ripalign}:} Interaction of two specific RNA
molecules, \emph{SL1} and \emph{SmY-10} of \emph{Caenorhabditis
elegans} as illustrated in Fig.~\ref{F:comversion}. The top 6
probable hybrids predicted by \texttt{rip} implemented by
\cite{rip2} is shown in column I. The hybrids listed in column II
are predicted by \texttt{ripalign} without any structure constraint.
The hybrids listed in Column III are predicted by \texttt{ripalign}
under the structural constraints that
$\textbf{5'-AAUUUUUG-3'}(R[56,62])$ and
$\textbf{3'-GUUUUAA-5'}(S[25,31])$ are Sm-binding sites (colored in
red) in \emph{SmY-10} and \emph{SL-1}, respectively. Here, we use
$J^{\sf {Hy}}_{i,j;h,l}$ to denote the hybrid induced by $R[i,j]$
and $S[h,l]$.}
\end{table}
\label{T:2}

{\bf (c): The \emph{U4}/\emph{U6} interaction}\\
Two of the snRNAs involved in pre-mRNA splicing, \emph{U4} and
\emph{U6}, are known to interact by base pairing
\citep{ZuckerAprison:88}. We divided all known metazoan \emph{U4}
and \emph{U6} snRNAs into three distinct groups and alignments:
protostomia without insects, insects and deuterostomia
\citep{Marz:08}. \cite{Marz:08} observed that insects behave in
their secondary structure different from other protostomes, see
Fig.~\ref{F:u4u6}. Comparing all the predicted \emph{U4}/\emph{U6}
interactions,
displayed in Fig.~\ref{F:u4u6}, we can conclude:\\
(1) the secondary partial structures of the \emph{U4}/\emph{U6}
complex for all three groups predicted by \texttt{ripalign} agree
predominantly with the described secondary structures in metazoans
\citep{Thomas:90,Otake:02,Shambaugh:94,Lopez:08, Shukla:02}, e.g.~as
depicted in Fig.~\ref{F:u4u6} (top) for \emph{C.~elegans}
\citep{ZuckerAprison:88}.\\
(2) for all three groups, Stem I and II (Fig.~\ref{F:u4u6}, top) are
highly conserved. External ascendancies, such as protein interactions
may stabilize stem II additionally.\\
(3) for all three groups, the $5'$ hairpin of \emph{U4} snRNA seems
highly conserved to interact with the \emph{U6} snRNA. This RNA
feature is not fully understood, since this element is also believed
to contain intraloop interactions and may bind to a 15.5kDa protein
\cite{Vidovic:00}.\\
(4) for all metazoans, the \emph{U6} snRNA shows conserved
intramolecular interactions between the $3'$ part  and the region
downstream of the $5'$-hairpin.\\
(5) for deuterostomes (Fig.~\ref{F:u4u6}, bottom), with a contact-region
probability of 45.5\%), our algorithm identifies a third \emph{U4}/\emph{U6}
interaction, Stem III, to be conserved, which agrees with the findings in
\cite{Jakab:97,Brow:95}. For protostomes, a similar feature with a
contact-region probability of $\leq 10\%$ can also be assumed.\\
(6) for both: protostomia (without insects) and deuterostomes, the
$5'$ hairpin of \emph{U6} snRNA seems to interact with the \emph{U4}
$3'$ hairpin. However, this observation does not hold for insects,
which agrees with a systematically different secondary structure of
spliceosomal RNAs in insects \citep{Marz:08}.\\
\begin{figure}
\begin{center}
\begin{tabular}{c}
  \cite{ZuckerAprison:88} \\
  \includegraphics[width=0.7\columnwidth]{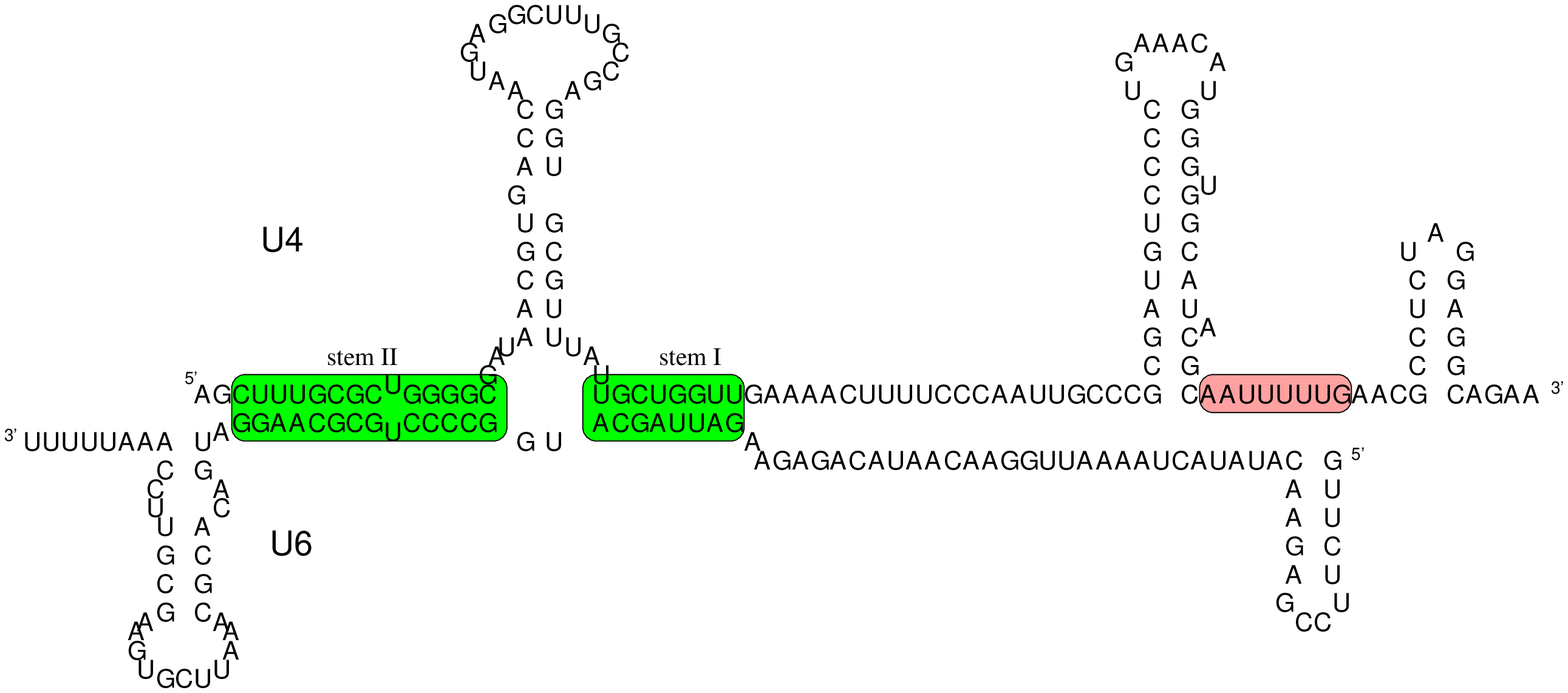}\\
  Protostomia without insects \\
  \includegraphics[angle=90,width=0.9\columnwidth]{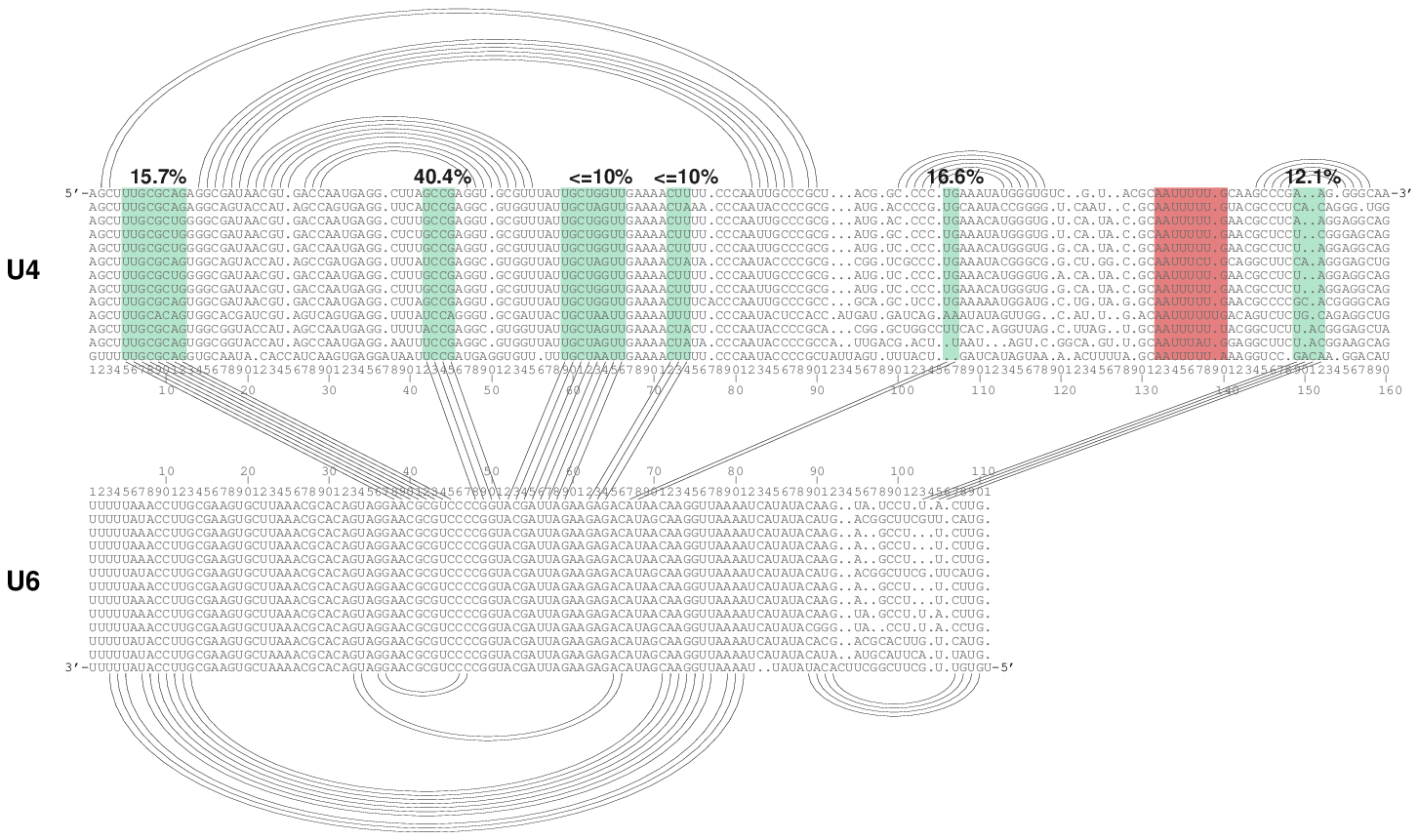} \\
  Insects \\
  \includegraphics[angle=90,width=0.9\columnwidth]{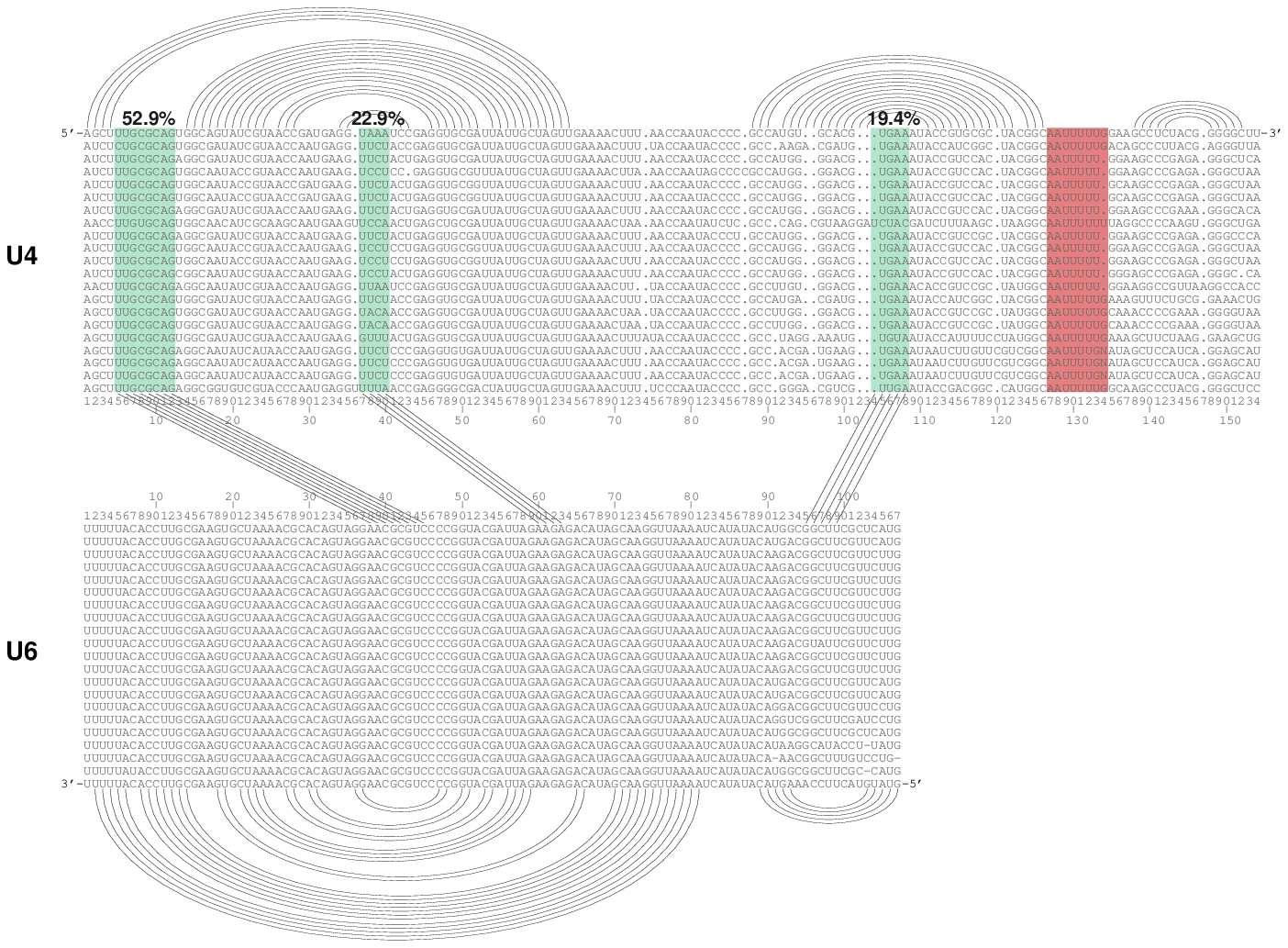} \\
  Deuterostomia \\
  \includegraphics[angle=90,width=0.9\columnwidth]{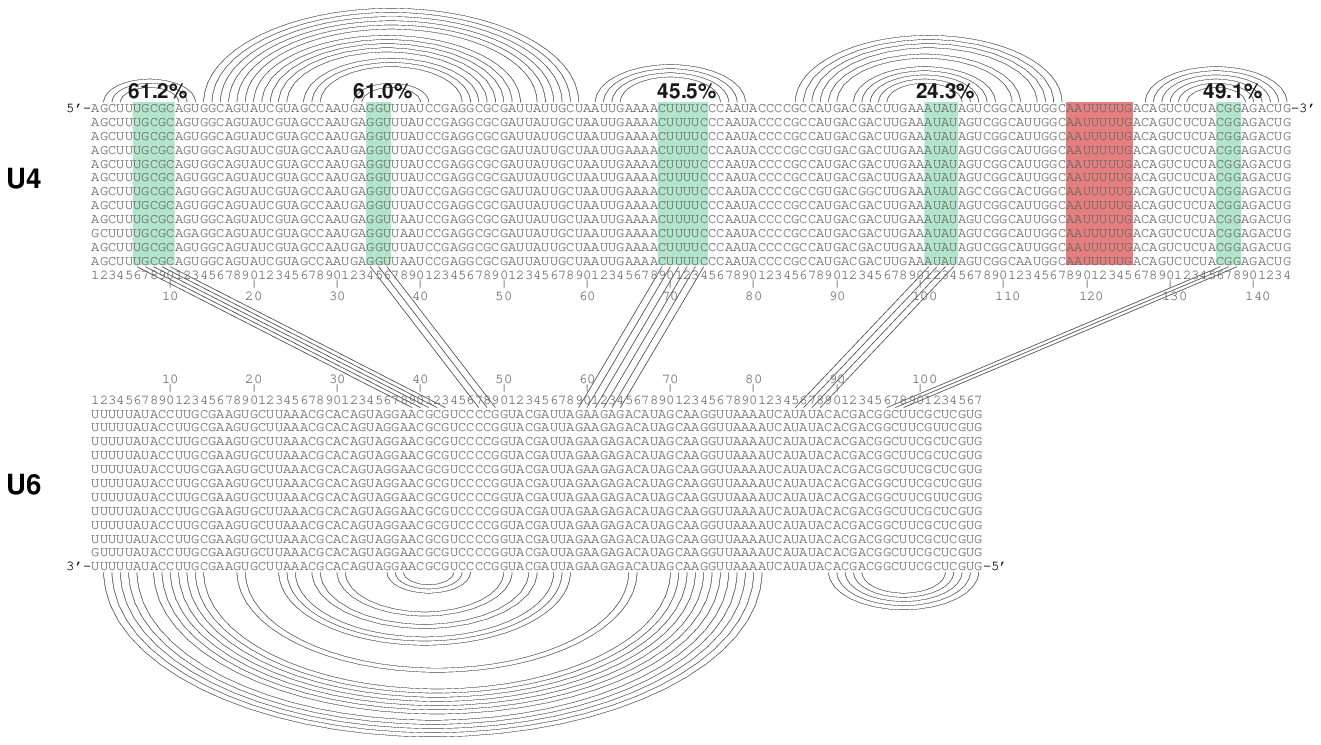} \\
\end{tabular}
\end{center}
\caption{\textbf{The \emph{U4}-\emph{U6} interaction prediction with
Sm-binding site constraint in \emph{U4}.} The Sm-binding site in
molecule \emph{U4} is \textbf{$5'$-AAUUUUUG-$3'$}(colored in red).
Top of the figure is the natural structure of \emph{U4}/\emph{U6} of
\emph{C. elegans} depicted by \cite{ZuckerAprison:88}, in which the
stem I, stem II and Sm-binding site are colored in green and red,
respectively. The joint structures of protostomia (without
insects), insects and deuterostomia (from top to bottom) are
predicted by \texttt{ripalign} under the Sm-binding site
constraint. The target site (green boxes) probabilities computed by
\texttt{ripalign} are annotated explicitly if $>10\%$ or just by
$\leq 10\%$, otherwise.} \label{F:u4u6}
\end{figure}

We finally remark that the quality of prediction of \texttt{ripalign}
depends critically on the quality of the MSAs.
This issue of alignment quality is not easily solved: creating an
alignment without knowing the structure is unlikely to produce a
structural alignment. It might be an option to realign the sequences
of an RNA family taking both the predicted secondary structures and
predicted joint structure with other RNA families into consideration.
Furthermore, \texttt{ripalign} is limited by its \emph{a priori}
output class of joint structures. Thus \texttt{ripalign} cannot identify
any joint structures exhibiting pseudoknots.
To save computational resources, we stipulate that only alignment
positions contribute as indices and loop sizes. The assumption may
cause, for instance, the existence of some interior arcs $R_{i}R_{j}$
having arc-length smaller than three.
\cite{Bernhart:08} showed that this problem can be improved
substantially by introducing a different, more rational handling of
alignment gaps, and by replacing the rather simplistic model of
covariance scoring with more sophisticated RIBOSUM-like scoring
matrices.

\end{methods}

\begin{methods}
\bigskip\par\noindent\textbf{Acknowledgements.}  We want to thank
Fenix W.D. Huang and Jan Engelhardt for helpful suggestions. We
thank Sharon Selzo of the Modular and BICoC Benchmark Center, IBM
and Kathy Tzeng of IBM Life Sciences Solutions Enablement. Their
support was vital for all computations presented here. We thank
Albrecht Bindereif, Elizabeth Chester and Stephen Rader for their
\emph{U4}/\emph{U6} analysis. This work was supported by the 973
Project of the Ministry of Science and Technology, the PCSIRT
Project of the Ministry of Education, and the National Science
Foundation of China to CMR and his lab, grant No.\ STA 850/7-1 of
the Deutsche Forschungsgemeinschaft under the auspices of SPP-1258
``Small Regulatory RNAs in Prokaryotes'', as well as the European
Community FP-6 project SYNLET (Contract Number 043312) to Peter F.
Stadler and his lab.
\end{methods}

\bibliographystyle{bioinformatics}
\bibliography{r6}
\end{document}